\documentclass{aa52}                      
\input{psfig.sty}                       
\begin{document}
%
%
%
   \title{ A Time Evolution Study of the Superhumps of the Dwarf Nova
   1RXS J232953.9+062814 by Wavelet Transform
   }

   \author{A.-Y. Zhou
             \inst{}
           \and Y.-L. Qiu
           }

   \offprints{A.-Y. Zhou}

   \institute{National Astronomical Observatories, Chinese Academy of Sciences,
             Beijing 100012, China\\
             \email{aiying@bao.ac.cn}
          }

   \date{Received ~~~~~~~~~~ 2002 / Accepted ~~~~~~~~~~ 2002}

   \abstract{
The time evolution behaviour of the superhumps of the dwarf nova
1RXS J232953.9+062814 is investigated with the wavelet analysis method.
On the basis of two nights CCD photometry performed during its first superoutburst
as well as other published brightness data,
we reveal the superhump's time-dependence as a function of periods and time.
Our light curves, which phased in the rapid decay ending portion of the superoutburst and
in the dawn of a following normal outburst, are important to help trace the superhump evolution for the star.
Evident amplitude variations of the superhumps, reflecting the fading of outbursts,
are detected.
The general profile of brightness fading over the outbursts
roughly followed an exponential decay law or a form of a five-order polynomial.
Both the superhump period and the orbital period of the binary system
are detected in the present data. We obtain
$P_{\rm sh}$=0.04575$\pm$0.00005\,d and $P_{\rm orb}$=0.04496$\pm$0.00005\,d.
They agree with the existing values based on additional data.
The two periods exchanged their roles during the superhump evolution.
   }

    \titlerunning{Superhump Evolution of 1RXS J232953.9+062814}
    \maketitle
%
%
\section{Introduction}
Cataclysmic variables (CVs) are close binary systems in which a low-mass secondary
transfers mass onto a white dwarf (Warner~\cite{warn95}), while dwarf novae are a sub-group of CVs
which experience repetitive outbursts with typical amplitudes of 2--5\,mag (Warner~\cite{warn87}).
The dwarf nova outbursts are considered to be a suddenly enhanced release of
gravitational energy induced by the thermal instability of an accretion dick (Osaki~\cite{osak96}).
During outbursts, short-period dwarf nova systems often show photometric oscillations
(superhumps) at periods of a few percent longer than their orbital periods ($P_{\rm orb}$).
The superhump period is thought to be the beat between $P_{\rm orb}$ and a tidally driven
precession of the eccentric disk.
According to the standard evolution model of compact binary systems
(Paczynski~\cite{pacz81}; King~\cite{king88}), orbital angular momentum losses maintain the mass-transfer
and thus drive the CV evolution.

1RXS J232953.9+062814 (equinox at 2001.11.03.926) was identified as
a cataclysmic variable by Wei et al.\ (\cite{wei99}).
It was classified as a dwarf nova by Hu et al.\ (\cite{hujy98}).
Its outburst was first detected on 2001 November 3.926 UT by P. Schmeer,
a member of the VSNET collaboration team\footnote{http://www.kusastro.kyoto-u.ac.jp/vsnet/}.
This team soon reported superhumps with amplitudes
of 0.2--0.3\,mag and a period of 0.046311$\pm$0.000012\,d, indicating that this object is a
dwarf nova of type SU Ursae Majoris (Uemura et al.\ \cite{uemu01}).
In terms of the quiescent spectra obtained by Hu et al.\ (\cite{hujy98}) and
Wei et al.\ (\cite{wei01}),
this object is a hydrogen-rich CV with a high-inclination accretion disc.
However, the observed short superhump period is below the
common `period minimum' for hydrogen-rich secondaries ($\sim$1.3\,h).
Further data showed that the star has a short distance of
140--350\,pc (Skillman et al.\ \cite{skil02}; Wei et al.\ \cite{wei01}) and
it is much brighter than V485 Centauri, the other known object of this class.
Therefore, the dwarf nova is a very important object for studying the evolutionary scenario of
CVs and it was immediately monitored by two research networks
after its detection of outburst with the aim to detect the time-evolution of
the superhumps and the system's properties (Skillman et al.\ \cite{skil02};
Thorstensen et al.\ \cite{thor02};
Uemura et al.\ \cite{uemu02}).
Our observations were carried out at the Xinglong Station of
the National Astronomical Observatories of China (NAOC).
This paper reports the results of a study on the superhump evolution based on
all available brightness data.

\section{Photometry }
\label{sect:Obs}
\begin{figure}[t]
   \begin{center}
   \vspace{2mm}
   \hspace{2mm}\psfig{figure=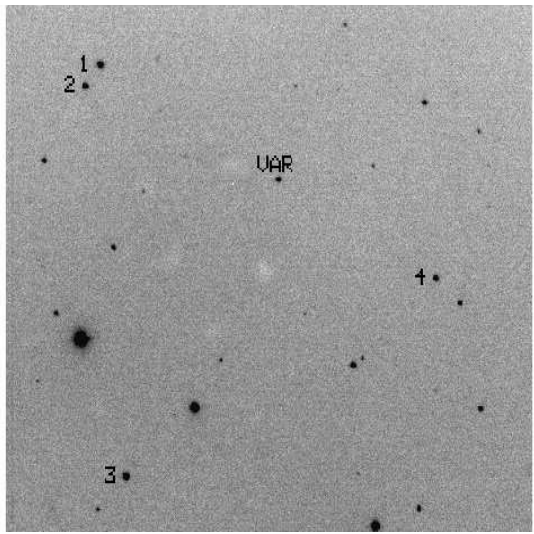,width=70mm,height=70mm,angle=0.0}
   \parbox{180mm}{{\vspace{2mm} }}
   \caption{ The CCD image of the filed of the dwarf nova 1RXS J232953.9+062814,
   obtained with the 60-cm supernova survey telescope at the Xinglong Station of NAOC.   }
   \label{Fig:finding-chart}
   \end{center}
\end{figure}

One night coordinated observations of the dwarf nova 1RXS J232953.9+062814 were
carried out on two telescopes at the Xinglong Station of NAOC on 2001 November 8.
The Johnson $V$ photometry was performed with the three-channel high-speed photoelectric
photometer, designed for the Whole Earth Telescope campaign
(Nather et al.~\cite{nath90}; Jiang \& Hu~\cite{jian98}),
mounted on the 85-cm Cassegrain telescope.
The star GSC 0591-1706
($\alpha_{2000}=23^{\rm h}30^{\rm m}11^{\rm s}.42$,
$\delta_{2000}=06^{\circ}26'07^{\prime\prime}.8$, $V=10.52$\,mag, K2);
was used as comparison as suggested by VSNET members. The dwarf nova,
comparison and sky background were simultaneously exposed in continuous 5-s intervals.
The unfiltered CCD photometry was carried out on the 60-cm telescope,
which dedicated to a supernova survey.
Another night CCD photometry was continued on 2001 November 9.
The CCD camera was made by Princeton Instruments, which is a liquid nitrogen cooled
CCD and it has a peak quantum efficiency of over 90\%.
It has 1340$\times$1300 pixels, each 20$\mu$m in size.
The field of view is $10'\times10'$
at the telescope's Cassegrain focus (Qiu et al.\ \cite{qiuy01}).
The integration time for each CCD frame is 10\,s.
In total, we obtained 3277 $V$ measurements together with 1785 CCD frames of
the dwarf nova field.
Figure~\ref{Fig:finding-chart} shows the CCD image of the field.
The standard CCD aperture photometry was made using the procedures implemented in the packages
of IRAF.
%

\begin{figure*}
   \begin{center}
   \vspace{2mm}
   \hspace{2mm}\psfig{figure=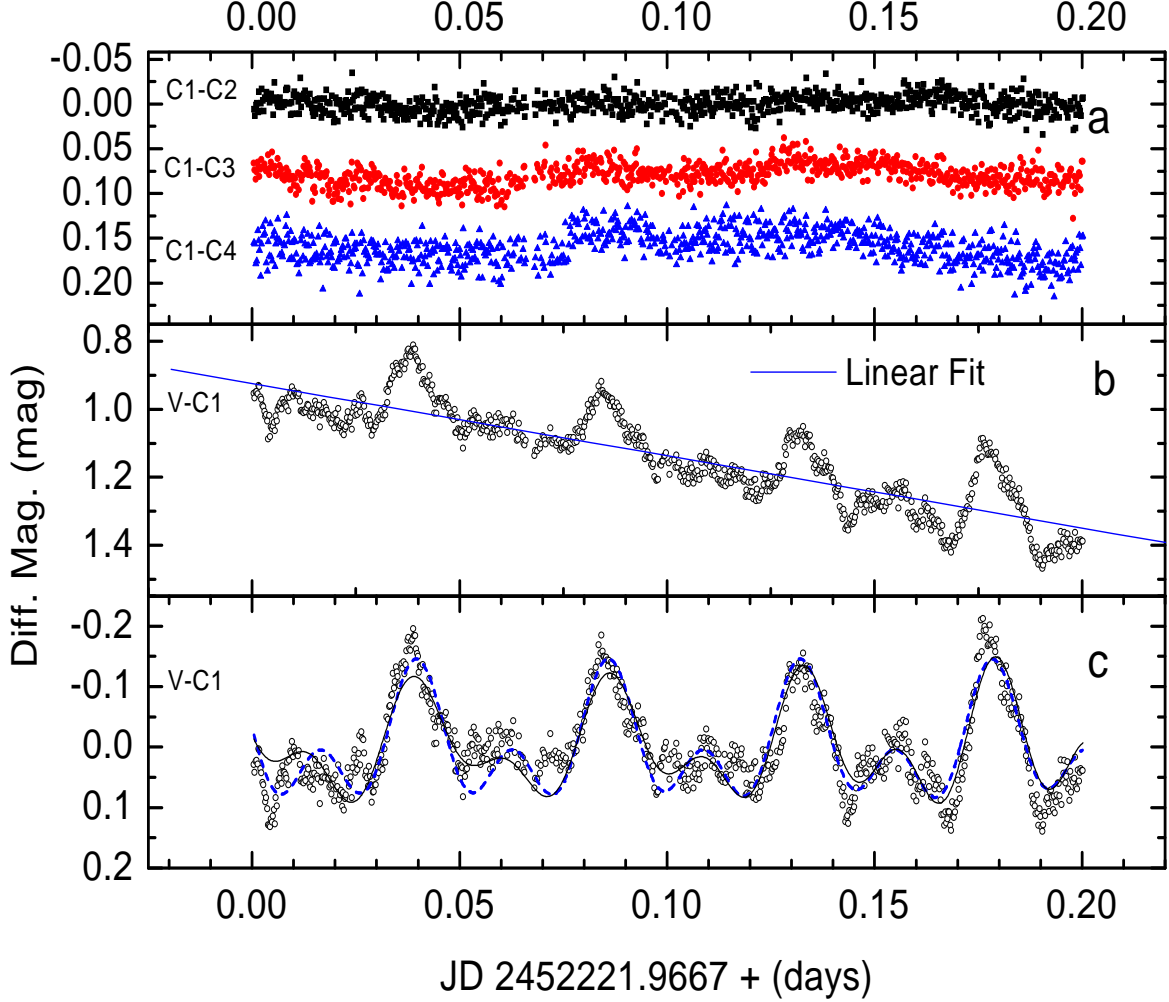,width=180mm,height=145mm,angle=0.0}
   \parbox{180mm}{{\vspace{2mm} }}
   \caption{ Light curves on 2001 November 8.
   (a): Differential magnitudes between the comparison stars C1, C2, C3 and C4.
   Means were subtracted.
   For display, the means of (C1--C3) and (C1--C4) were shifted to 0.08 and 0.16\,mag
   from zero, respectively.
   (b): Differential magnitudes of 1RXS J232953.9+062814 relative to C1, (V--C1),
   and a linear fit: $0.924\pm0.005 + (2.125\pm0.044)~t, t=$JD--2452221.9667\,d.
   (c): (V--C1) but the linear trend was removed, along with sinusoids of periods
   $P_{\rm sh}$ (plus 0.5$P_{\rm sh}$, dashed line) and of both $P_{\rm sh}$ and
   $P_{\rm orb}$ (solid line). }
   \label{Fig:lightcurve-1108}
   \end{center}
\end{figure*}

\begin{figure*}
   \begin{center}
   \vspace{2mm}
   \hspace{2mm}\psfig{figure=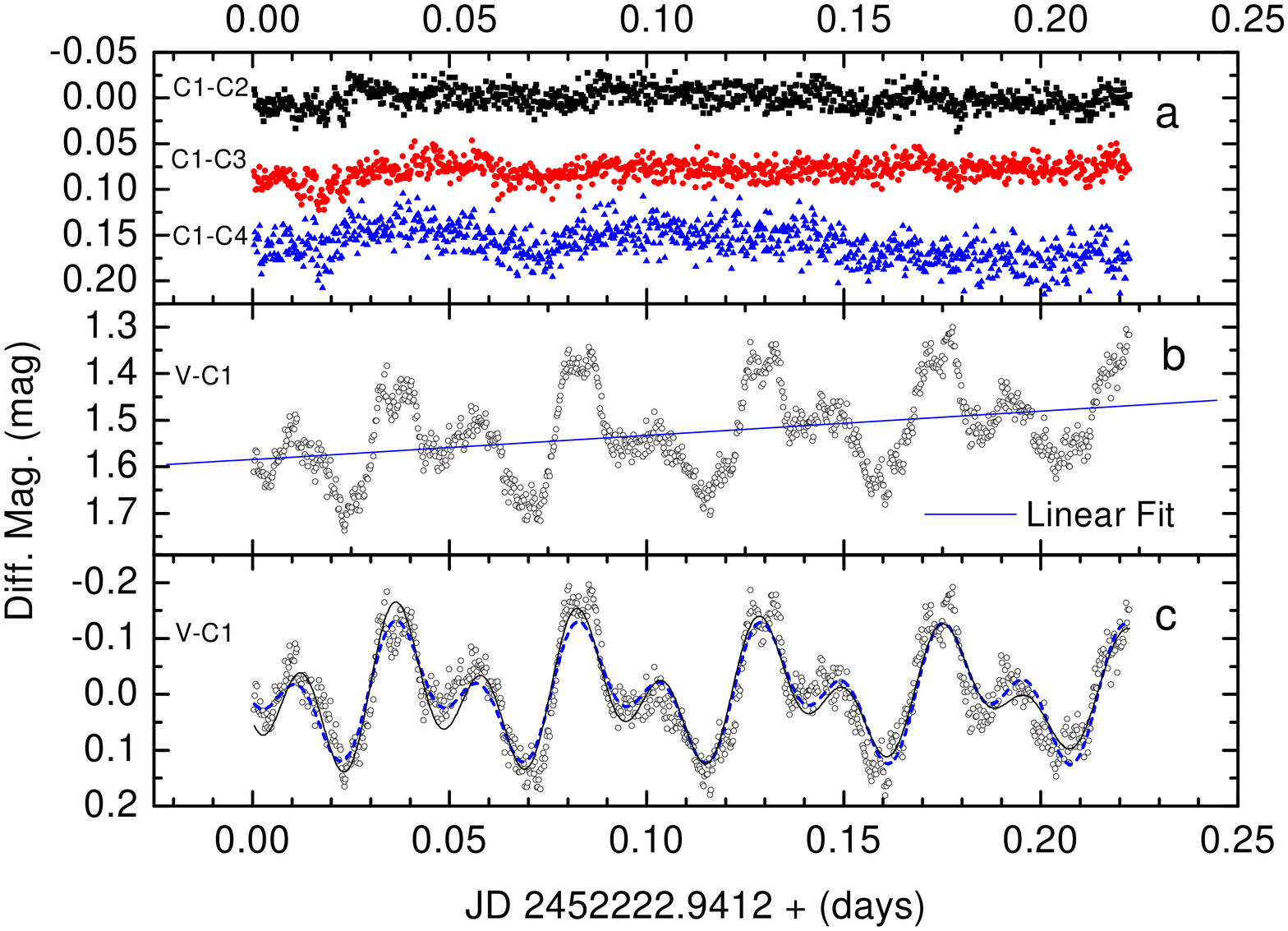,width=180mm,height=145mm,angle=0.0}
   \parbox{180mm}{{\vspace{2mm} }}
   \caption{ Light curves on 2001 November 9.
   (a): Differential magnitudes between the comparison stars C1, C2, C3 and C4.
   Means were subtracted.
   For display, the means of (C1--C3) and (C1--C4) were shifted to 0.08 and 0.16\,mag
   from zero, respectively.
   (b): Differential magnitudes of RX 2329+06 relative to C1, (V--C1),
   and a linear fit: $1.585\pm0.006 - (0.520\pm0.043)~t, t=$JD--2452222.9412\,d.
   (c): (V--C1) but the linear trend was removed, along with sinusoids of periods
   $P_{\rm sh}$ (plus 0.5$P_{\rm sh}$, dashed line) and of both $P_{\rm sh}$ and
   $P_{\rm orb}$ (solid line). }
   \label{Fig:lightcurve-1109}
   \end{center}
\end{figure*}

Unfiltered CCD differential magnitudes were established with respect to an assemble comparison
star consisting of several stars (C1=GSC 0591-1660, 14.1 mag;
C2=GSC 0591-1692, 14.4 mag; C3=GSC 0591-1665, 14.0;
C4: $\alpha_{2000}=23^{\rm h}29^{\rm m}42^{\rm s}.31$,
$\delta_{2000}=06^{\circ}26'10^{\prime\prime}.2$, USNO, 14.5R) in the observed
field (Fig.\ref{Fig:finding-chart}).
The typical observational accuracy for the CCD data is $\sim$0.01\,mag.
From panel `a' of Fig.~\ref{Fig:lightcurve-1108},
the standard deviations for the three sets of differential CCD measurements (C1--C2),
(C1--C3) and (C1--C4) are 0.010, 0.013 and 0.018\,mag, respectively.
The light curves are presented in Figs.~\ref{Fig:lightcurve-1108} and~\ref{Fig:lightcurve-1109},
where the fits (solid and dashed lines) with the superhump period and orbital period
are also plotted.

The photoelectric data, unfortunately, had some problems.
It seemed that the sky and comparison channels were abnormal during the observations.
We failed to use them to establish differential magnitudes for the variable.
Finally, we used the variable's measurements alone after subtracting a second-order
polynomial fit. The results are only valuable for a rough estimate of the light variations
in the star. The light curves have been given in Wei et al.\ (\cite{wei01}).
For this reason, the data were not considered in the following analyses.

%

%

%

\section{Wavelet analysis}
\label{sect:analysis}
In the analysis of time-series data of variable stars, beyond the detection
of periodicities, one needs to investigate the stability of
the periodicities as the oscillations are caused by matter transfer
from a component to the other as in CVs.
The detection of a variable period is very important as it gives information
on the binary nature or on the evolutionary status of a star.
However, it is difficult to determine the period variability. The conventional
Fourier analysis is unable to give any information about period variation
due to it just localizes frequency information without time localization.
Unlike Fourier analysis, in which we analyze signals using sines and cosines,
alternatively, wavelet analysis uses wavelet basis to decompose and
reconstruct a periodic signal or an oscillation.  The wavelet basis localizes
both in time and frequency spaces, so the wavelet analysis is a well suited
method to accomplish the mission of the detection of time dependence of oscillation frequencies.
In fact, it has been proving to be a versatile tool to disclose periodic and time-dependent
behaviour of a signal.
Moreover, wavelet transforms do not ask a long-time coverage data set as Fourier transforms do
in resolving periodicities in a signal.

There are a quite number of examples using wavelet analysis method in the study of
various oscillating behaviours. For instance, Goupil et al.\ (\cite{goup91}),
Szatm\'{a}ry \& Vink\'{o}~(\cite{szat92}), Scargle et al.\ (\cite{scar93}),
Coupinot et al.\ (\cite{coup92}) and Starck et al.\ (\cite{star97}).
In special,  Fritz \& Bruch (\cite{frit98}) used wavelet transforms to
study the flickering light curves in CVs.
The applicability of wavelet analysis for studying periodicity in the
light curves of variables has been further explored by some astronomers such as
Szatm\'{a}ry et al.\ (\cite{szat94}, \cite{szat96}) and Foster (\cite{fost96}).
The wavelet transform,
as these authors have demonstrated, is well-suited to detect the local
behaviour of the light curves of variables. For example, the investigation
of time-dependent phenomena including amplitude and frequency modulation,
changes of period and phase.
In view of the unevenly spaced data points, the ability of several improved
wavelet transforms to detect and quantify periodic and pseudo-periodic signals,
the weighted wavelet Z-transform (WWZ) developed by Foster (\cite{fost96}) was used.

\begin{figure}[t]
   \begin{center}
   \vspace{2mm}
   \hspace{1mm}\psfig{figure=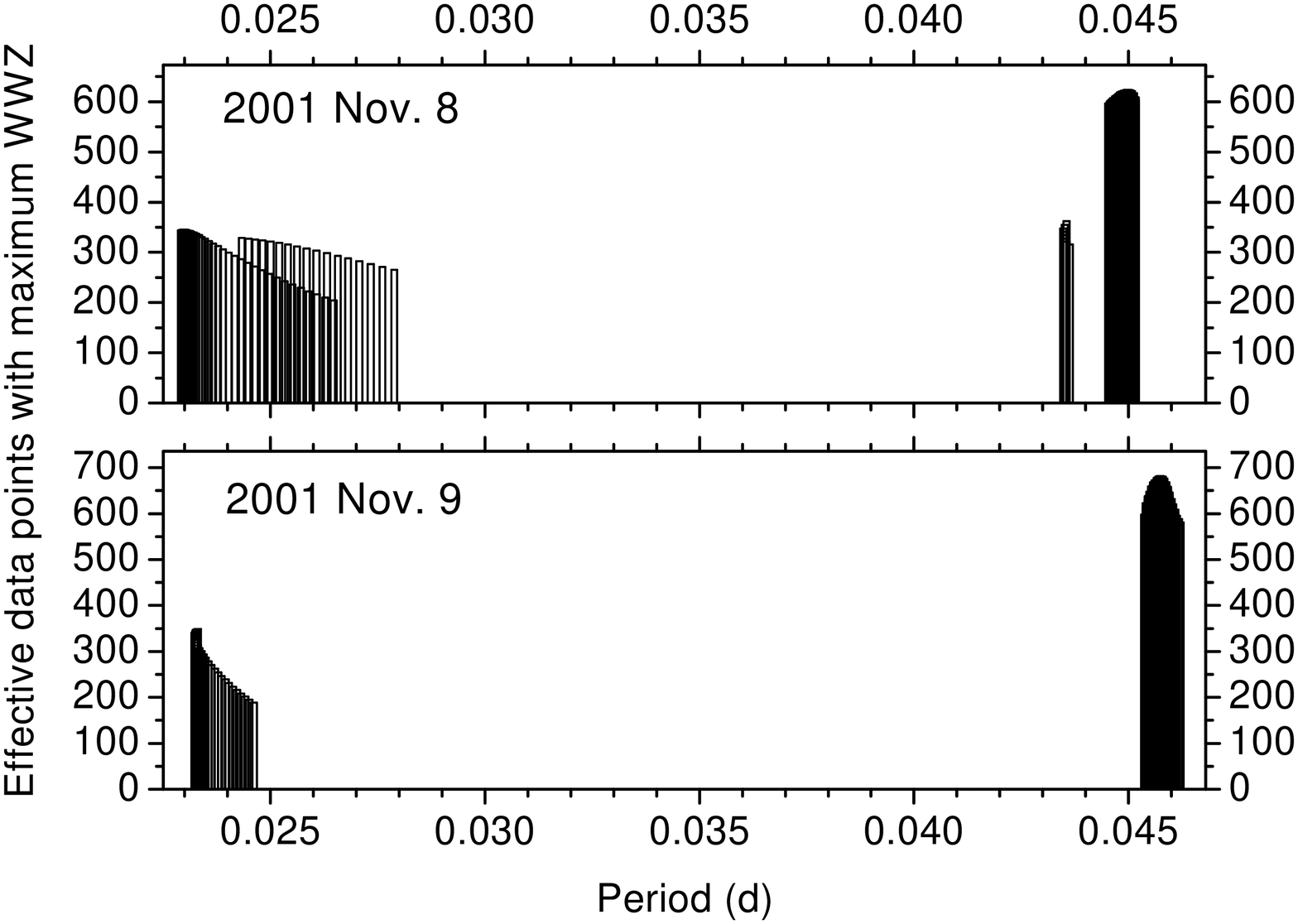,width=85mm,height=85mm,angle=0.0}
   \caption{ The distribution of the numbers of effective data points
   at which WWZ reaches maxima and we regard the corresponding periods
   as the main contents of the light variations at a time.  }
   \label{Fig:wwz-Neff}
   \end{center}
\end{figure}

We set trial detection for frequencies from 8 to 50 cycle d$^{-1}$
(corresponding to periods in the range of 0.02 to 0.125\,d). Times varied from 0 to 0.2\,d
in a step of 0.002\,d. Frequency resolution (or calculation error) was 1.0 cycle d$^{-1}$.
This tended to establish rough limits for the fluctuation frequency as a function of time.
We obtained frequencies at 23, 36, 37, 38, 39, 40, 41, 22, 44, 43, 42 and 41 cycle d$^{-1}$
in order with time increase for the 2001 Nov. 8 data,
at which the WWZ transform reaches maxima.
As for the 2001 Nov. 9 data, we obtained 41, 42, 43 and 22 cycle d$^{-1}$.
Most of the data points at most times were of frequency 22 cycle d$^{-1}$,
a value close to what we want to find, the superhump period 0.0463\,d (21.59 cycle d$^{-1}$) and
the orbital period 0.0446\d (22.42 cycle d$^{-1}$), respectively (see
the distribution in Fig.~\ref{Fig:wwz-Neff}).
At the first place, we see the frequency or period was not constant
during the observing span. In addition, the two sets of trial periods appeared in opposite phases.
This is why we use wavelet analysis rather than Fourier analysis.
Fourier transform always assumes periods to be constant in the investigated duration,
so it prevents us from uncovering the variability of a period.
Then we refined the step-length of frequency to be 0.02 cycle d$^{-1}$,
corresponding to an error of $\sim$0.00005\,d in period for the star,
to obtain more accurate values of periods.
We inferred that the orbital period (0.0446 d, frequency 22.42 cycle d$^{-1}$) or
its daily alias (21.42) and harmonic (44.84) were mainly present in the first night data,
while the superhump period (0.0463, frequency 21.60 cycle d$^{-1}$) or
its daily alias (22.60) and harmonic (43.20) were mainly present in the second night data.
No trace of 22.42 (orbital frequency) was found in the second data set.
The second night data mainly reflect the superhump content oscillating at 21.59.
Finally, according to the distribution of periods and numbers of data points
where WWZ values reached maxima (Fig.~\ref{Fig:wwz-Neff}),
we obtained frequencies at 22.24$\pm$0.02 cycle d$^{-1}$ ($P_{\rm orb}$=0.04496$\pm$0.00005\,d)
representing the orbital motion of the binary system from the 2001 Nov. 8 data,
and at 21.86\,cycle d$^{-1}$ ($P_{\rm sh}$=0.04575$\pm$0.00005\,d),
the photometric oscillations of the dwarf nova.
The results agree with the values 0.0446\,d (64.177 m) and
0.04637(4)\,d (66.672 m), respectively, given by Skillman et al.\ (\cite{skil02}) and
Uemura et al.\ (\cite{uemu02}).
%

\begin{figure}[b]
   \begin{center}
   \vspace{39mm}
   \hspace{1mm}\psfig{figure=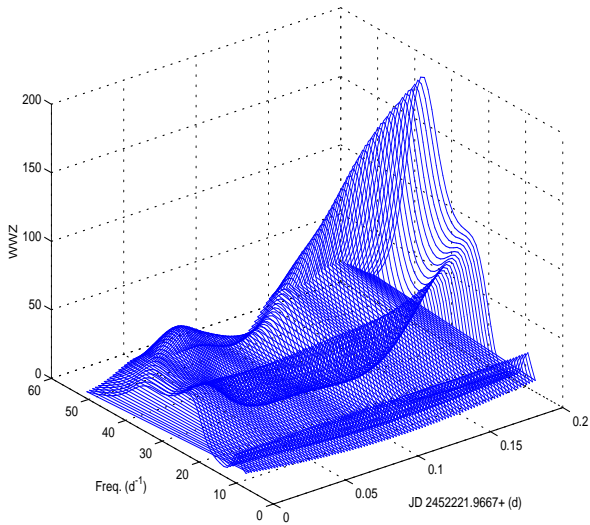,width=80mm,height=70mm,angle=0.0}
   \caption{The wavelet solution of the superoutburst on 2001 Nov. 8.
   It shows the time evolution of the superhumps as a function of time and frequency (period) }
   \label{Fig:wavelet-1108}
   \end{center}
\end{figure}

\begin{figure}[t]
   \begin{center}
   \vspace{17mm}
   \hspace{1mm}\psfig{figure=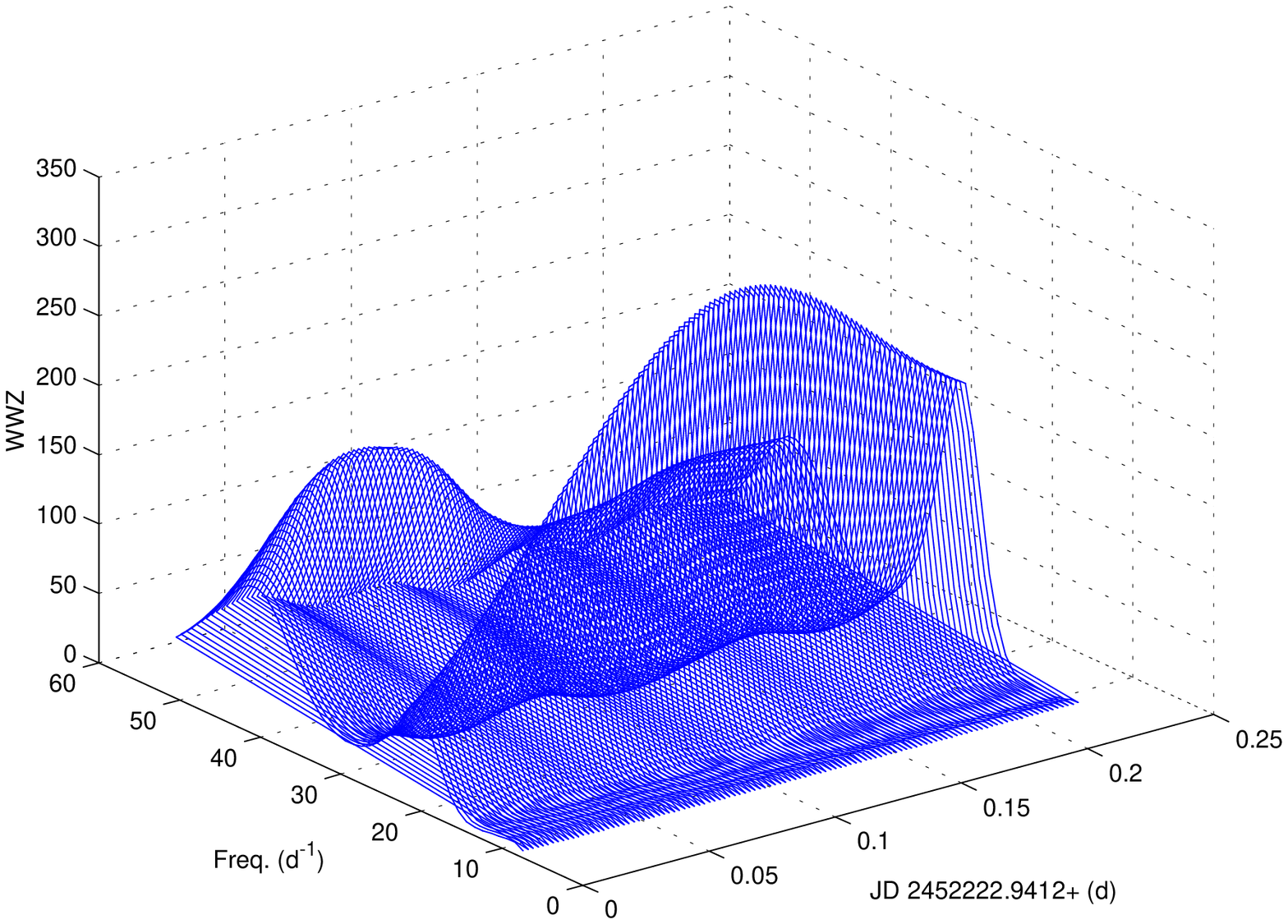,width=80mm,height=70mm,angle=0.0}
   \caption{The wavelet solution of the superoutburst on 2001 Nov. 9.
   It shows the time evolution of the superhumps as a function of time and frequency (period). }
   \label{Fig:wavelet-1109}
   \end{center}
\end{figure}

\begin{figure}[]
   \begin{center}
   \vspace{2mm}
   \hspace{1mm}\psfig{figure=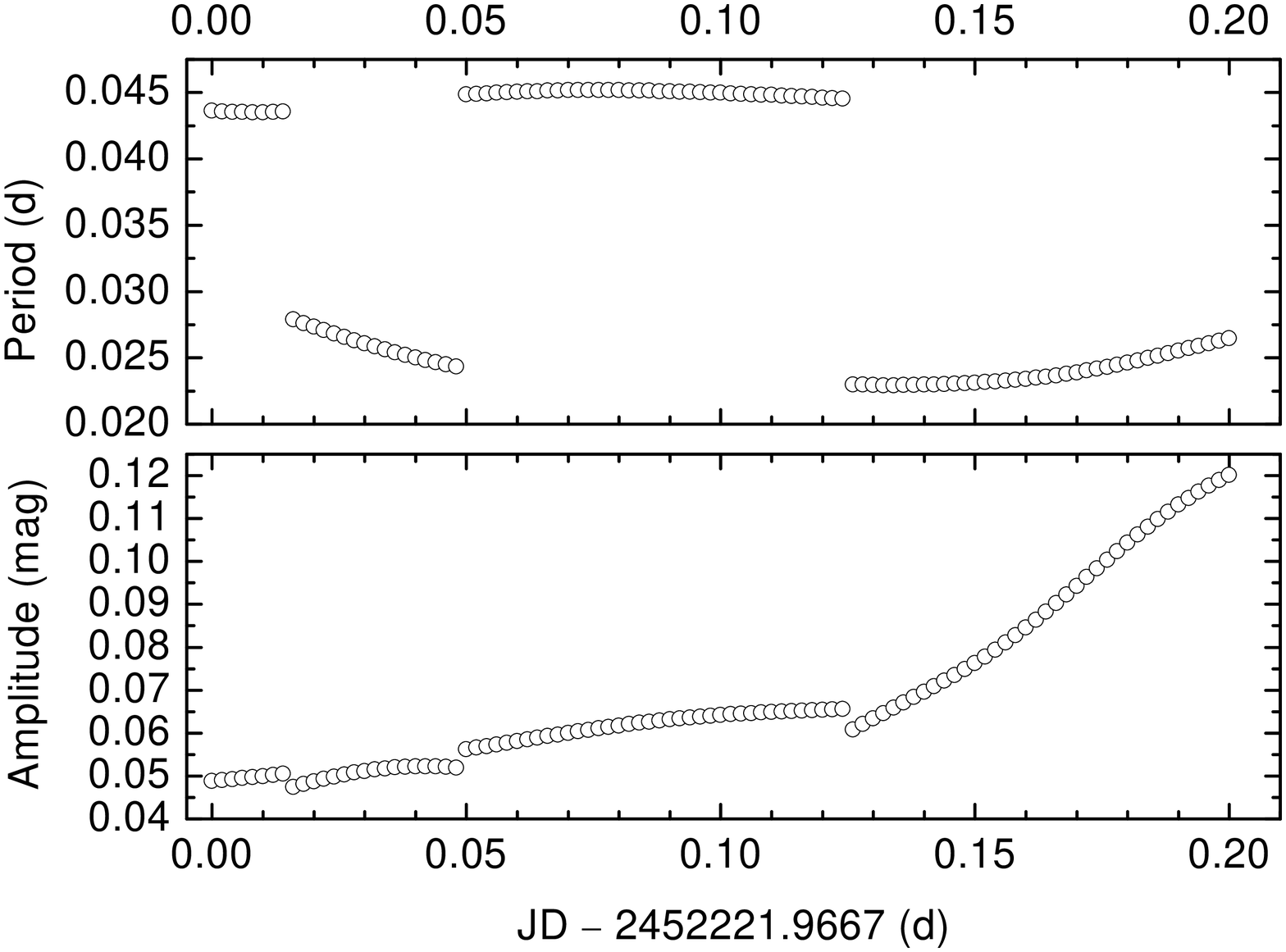,width=85mm,height=95mm,angle=0.0}
   \caption{ Time evolution of period, amplitude as determined by wavelet transform
   for the data on 2001 Nov. 8.   }
   \label{Fig:per-ampl-1108}
   \end{center}
\end{figure}

\begin{figure}[]
   \begin{center}
   \vspace{2mm}
   \hspace{1mm}\psfig{figure=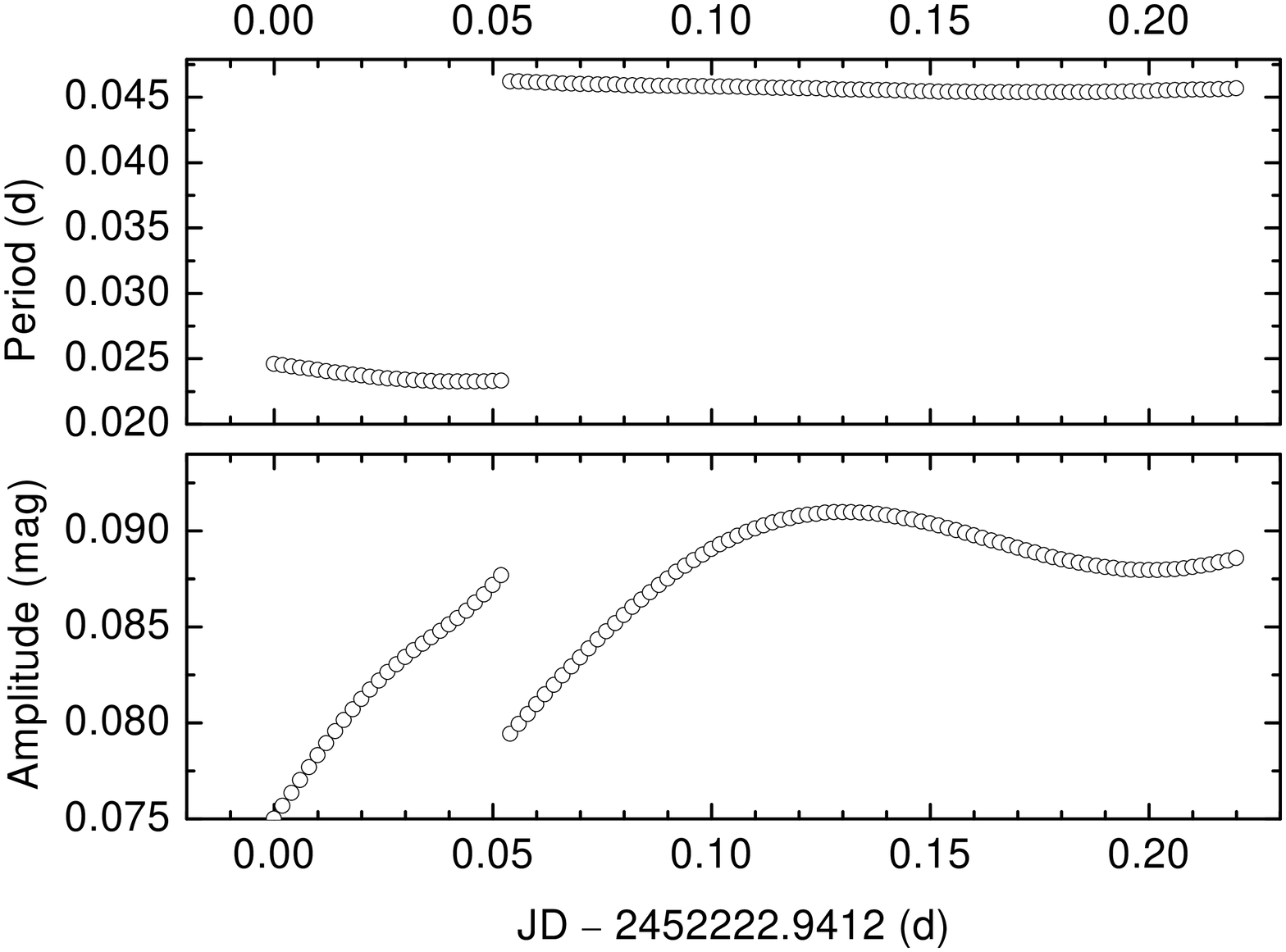,width=85mm,height=95mm,angle=0.0}
   \caption{ Time evolution of period, amplitude as determined by wavelet transform
   for the data on 2001 Nov. 9.   }
   \label{Fig:per-ampl-1109}
   \end{center}
\end{figure}

The results of wavelet analyses including the variation of superhumps as
a function of both time and frequency, the time-evolution of amplitudes and periods
of the superhumps are presented in Figs.~\ref{Fig:wavelet-1108} to \ref{Fig:per-ampl-1109}.
We used the raw light curves (see panels `b' of Figs.~\ref{Fig:lightcurve-1108} and
~\ref{Fig:lightcurve-1109}) without linear-fits removed in the analyses.
This ensures all information in the brightness was not lost.
%

%
\clearpage
%
%

\section{Results and Discussion}
\label{sect:results}
\begin{figure}
   \begin{center}
   \vspace{2mm}
   \hspace{-2mm}\psfig{figure=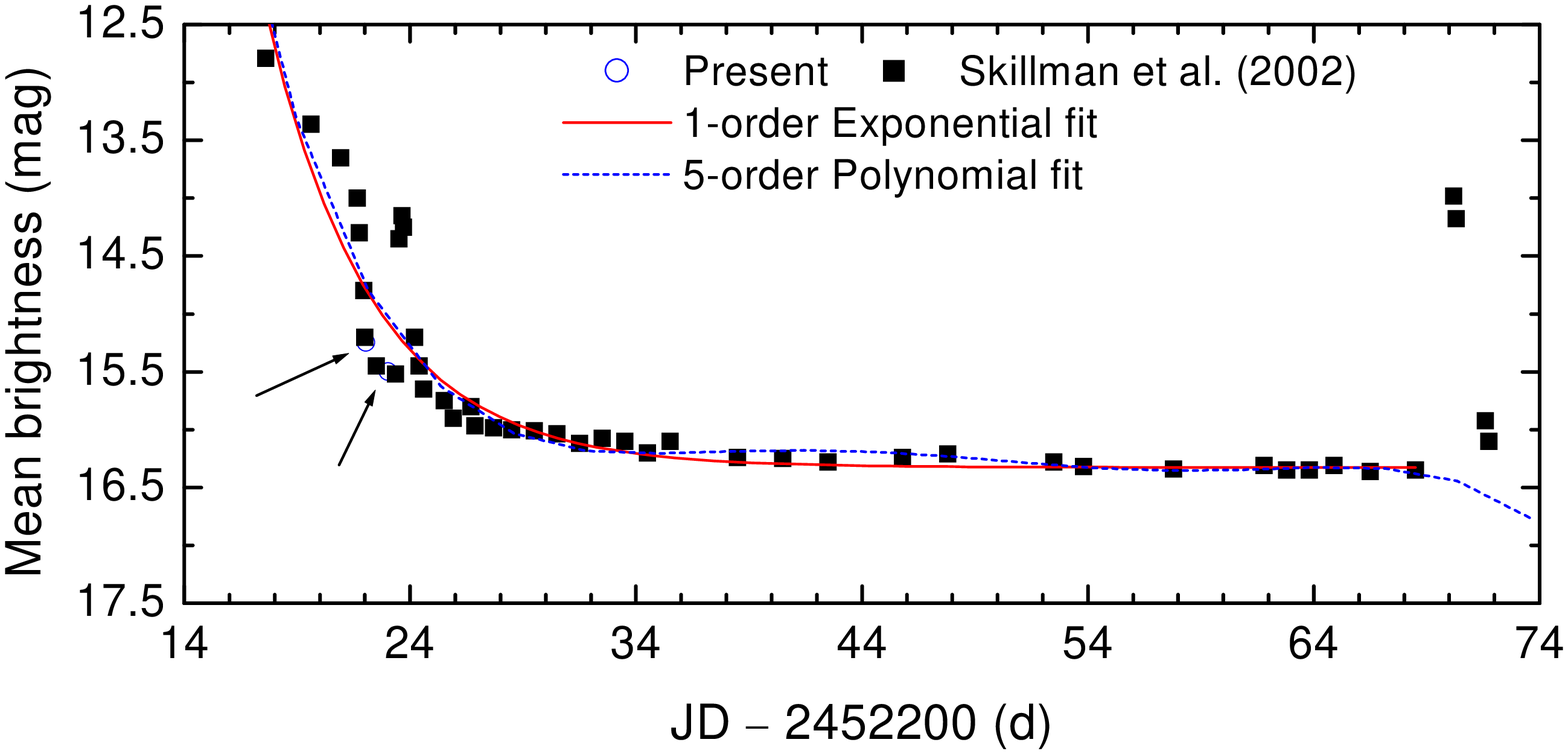,width=85mm,height=75mm,angle=0.0}
   \caption{ Mean light curves of 1RXS J232953.9+062814 during the 2001 outbursts.
   This is a reproduction of fig.~1 of Skillman et al.\ (2002) together with our data (two dots,
   indicated by arrows) and two decay fits.  }
   \label{Fig:hump-evol}
   \end{center}
\end{figure}

We analysed the superhump evolution in the interacting binary system 1RXS J232953.9+062814 based on
the two nights data (2001 Nov. 08--09) obtained by speed CCD photometry.
Wavelet analyses reveal that the periodicities in the light curves changed with time
(see the top panels of Figs.~\ref{Fig:per-ampl-1108} and~\ref{Fig:per-ampl-1109})
because the exchange of the roles of the superhump and orbital motion.
Superhump is prominent during outbursts while orbital motion of the binary system is conspicuous
in quiescence. In our case, superhump period is noticeable on 2001 Nov. 9, when the star
went into a new normal outburst, while the orbital period is remarkable on 2001 Nov. 8,
when the superoutburst evolved to its ending portion of fading.
According to the Fourier solution of the light curves by
Skillman et al.\ (\cite{skil02}) and the solution using Phases Dispersion Minimization by
Uemura et al.\ (\cite{uemu02}),
the periodicities involved in the superoutburst are mainly contributed by the superhumps,
while in quiescence the light variation mainly reflects the orbital motion of the binary system.
In other words, superhump period dominates the superoutburst while orbital period
contribution governs the quiescence.
Therefore, it should be that there is a time containing more information on the two kinds
of variations (orbital and oscillating). This time is now clearly between our observations,
i.e. between JD 2452222.0 and 2452223.0.
Our observations fell into the phase near the tail part of rapid decline of the superoutburst
and in sight of the first subsequent normal outburst.
Figure~\ref{Fig:hump-evol} shows the present data in dots (pointed by arrows). This figure was
reproduced from fig.~1 of Skillman et al.\ (\cite{skil02}), which can also be referred to fig.~1
of Uemura et al.\ (\cite{uemu02}).
Furthermore, from the middle panels of
Figs.~\ref{Fig:lightcurve-1108} and~\ref{Fig:lightcurve-1109}, it is obvious that
the brightness evolved in opposite directions over the two nights,
i.e. on 2001 November 8 the star faded gradually
but on November 9 it turned to be brighter and brighter.
In addition to the overall profile of the mean brightness changes,
we have noticed the linear decay trends involved in the nightly light curves.
The linear changing contents for these two sets of data:
$0.924\pm0.005 + (2.125\pm0.044) t, t=$JD--2452221.9667 d and
$1.585\pm0.006 - (0.520\pm0.043) t, t=$JD--2452222.9412 d clearly show that
our observations were in the transition phase of the superoutburst.
The change rate for these two nights are different: the former is faster than the latter.
The fact is that the star really run into its second brief outbursts (or an `echo outburst')
from JD 2452223.0 d, i.e. 2001 November 9, UT 12:00 (see
fig.~1 of Skillman et al.\ \cite{skil02}).
This indicates the value of the present data for tracing the star's outbursts evolution.
It also explains the two known periods resolved by the wavelet analyses based merely on our data.

Our high-speed photometry shows the detailed light variations of the superoutburst of
dwarf nova 1RXS J232953.9+062814, such as the linear decay content of brightness,
the superhumps and the doubled-humped wave of orbital motion.
Because both the superhump period and orbital period are involved in the light curves,
either the oscillation or orbital motion cannot reproduce the observed light curves well.
In Figs.~\ref{Fig:lightcurve-1108} and~\ref{Fig:lightcurve-1109}, the dashed curves are
the sinusoids of periods $P_{\rm sh}$=0.0463\,d (superhump) and 0.5$P_{\rm sh}$,
solid curves are of periods $P_{\rm sh}$, 0.5$P_{\rm sh}$ plus $P_{\rm orb}$=0.0446\,d (orbital).

Wavelet analyses also reveal the details of the fading of the outbursts.
The bottom panels of Figs.~\ref{Fig:per-ampl-1108} and~\ref{Fig:per-ampl-1109} display
the changes of the amplitudes of corresponding periods.

We may get an insight into the brightness changes in a general point of view.
Therefore, we made a first-order exponential decay fit and
a five-order polynomial fit to all the mean light curves available
(fig.~1 of Skillman et al.\ \cite{skil02})
with the exception of the two normal outbursts.
The results indicate that the brightness decayed following an exponential law
of $y (\rm mag.) = 16.326\pm0.064 - (168.34\pm60.45) e^{-{\it t}/(4.693\pm0.408)}$ or
in a five-order polynomial form of
$y = -30.570\pm8.706 + (5.110\pm1.205)t - (0.219\pm0.064)t^2 + (0.00457\pm0.0016)t^3 -
(5\pm2)\times10^{-5}t^4 + (1.87\pm0.91)\times10^{-7}t^5 $. Here $t$= JD--2452200 (d).
They are displayed in Fig.~\ref{Fig:hump-evol}, where the dashed line for polynomial fit
while the solid line for exponential fit.

%

\begin{acknowledgements}
Wavelet analysis was performed using the computer program WWZ,
developed by the American Association of Variable Star Observers.
We would like to thank Dr. Taichi Kato at the Univ. of Kyoto, Japan,
who suggested this observation.
This work was funded by the National Natural Science Foundation of China.
\end{acknowledgements}

\label{lastpage}

\end{document}